\documentclass[fleqn,twoside]{article}
\usepackage{espcrc2}
\usepackage{epsf}
\usepackage{graphicx}
\usepackage{epsfig}
\usepackage[fiugreright]{rotating}

\newcommand{\be}{\begin{eqnarray}}

\newcommand{\ee}{\end{eqnarray}}

\newcommand{\AmS}{{\protect\the\textfont2
  A\kern-.1667em\lower.5ex\hbox{M}\kern-.125emS}}

\title{{The Phase Diagram of QCD and Some Issues of   Large $N_c$ }\footnote{Talk prepared for a round table discussion at Fundamental Challenges in QCD, the 47'th International Unversitatswochen fur Theoretische Physics, Schladming, Styria, Autria, Feb. 2009.}}

\author{Larry McLerran\\
{\small\it Physics Department and Riken Brookhaven Center}\\ 
{\small\it PO Box 5000,  Brookhaven National Laboratory,  Upton, NY 11973 USA}\\
}

\begin{document}

\begin{abstract}
The large $N_c$ limit provides a good phenomenology of meson spectra and interactions.
I discuss some problems with applying the large $N_c$ approximation to the description of baryons,
and point out a number of apparent paradoxes and phenomenological difficulties
\vspace{1pc}
\end{abstract}
\maketitle

\section{Introduction}

\begin{figure}[!htb]
\begin{center}
\includegraphics[width=50mm,angle=270]{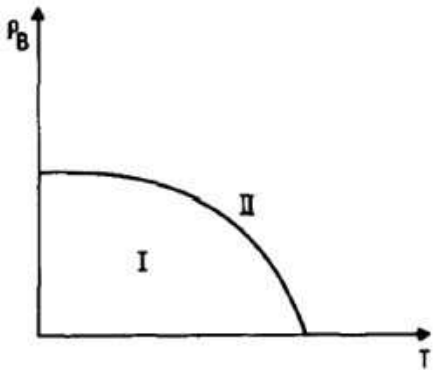}
             \end{center}
\caption[*]{\small  The phase diagram of Cabibbo and Parisi.
 }
     \label{phasediagram}
\end{figure}

\begin{figure}[htb]
\begin{center}
  \mbox{{\epsfig{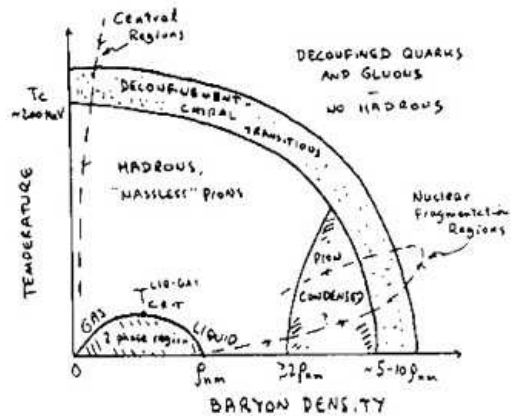}}
           }
   \end{center}
\caption[*]{\small The phase diagram as envisioned by Gordon Baym at the 1983 Long Range Plan.}
     \label{baym}
\end{figure}

The phase diagram of QCD at finite baryon density and temperature was to my knowledge first envisioned by Cabiibo and Parisi,\cite{Cabibbo:1975ig} and is shown in Fig. \ref{phasediagram}.
By 1983,  Gordon Baym had drawn the phase diagram shown in Fig. \ref{baym},
a phase diagram that had become more sophisticated and included the possibility of a liquid-gas phase transition at relatively low temperature and density, and pion condensation at intermediate density.

\begin{figure}[!htb]
\begin{center}
  \mbox{{\epsfig{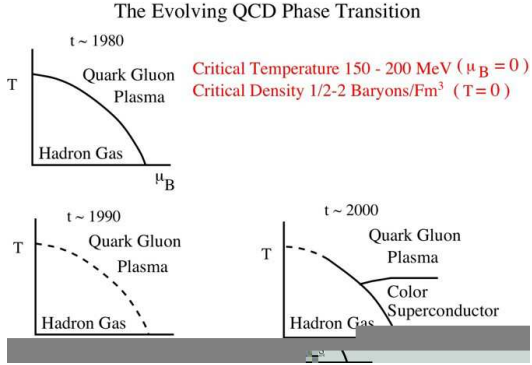}}
           }
   \end{center}
\caption[*]{\small (a) The time evolution of the phase diagram of QCD.}
     \label{mclerran_pd}
\end{figure}

In 2003, I presented the phase diagram as a function of time, Fig. \ref{mclerran_pd}, showing how our knowledge of the properties of high energy density matter has evolved over the years.\cite{McLerran:2002jb}
This evolution involved changing in our thinking about the nature of the transition between 1980-1990,
from a first order phase transition between confined matter and a quark gluon plasma, to that of a continuous cross over.  Between 1990 and 2000, the possibility of a Color Superconducting phase became compelling.\cite{Rapp:1997zu},\cite{Alford:1997zt} , and arguments were presented
that there is a first order phase transition at finite density and zero temperature, which evolves towards a critical endpoint as the temperature increases and then becomes a cross over.\cite{Alford:1997zt}  About the only thing which did not change much on the phase diagram were the temperatures and densities characterizing the transition between the confined phase and the quark gluon plasma.

During the past ten years, the richness of the structure of matter at very high energy density has been
appreciated:  The matter in the initial states of the colliding nuclei is a high density, highly coherent collection of gluons, the Color Glass Condensate.\cite{McLerran:1993ni},\cite{McLerran:1993ka},\cite{Iancu:2000hn},\cite{Iancu:2001ad},\cite{Ferreiro:2001qy}  The matter produced immediately after the collisions is composed of highly coherent gluons fields of very high energy density, the Glasma.\cite{Kovner:1995ja},\cite{Kovner:1995ts},\cite{Krasnitz:1999wc},\cite{Krasnitz:1998ns},\cite{Lappi:2003bi}  Such forms of matter cannot be represented on a phase diagram involving finite temperature and density, since the parameters which describe such matter are neither the temperature nor a conserved baryon number density.
\begin{figure}[!htb]
\begin{center}
  \mbox{{\epsfig{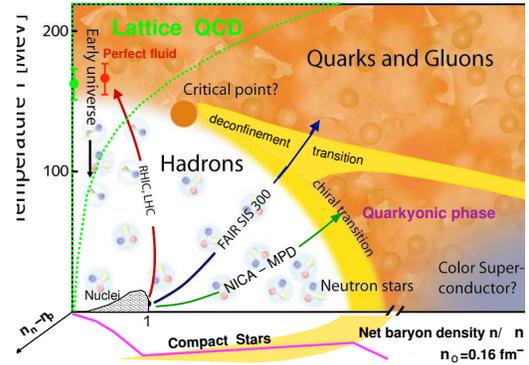}}
           }
   \end{center}
\caption[*]{\small The  current wisdom on the phase diagram of nuclear matter. }
     \label{blaschke}
\end{figure}

In the past few years, it has been argued that there is another phase of matter at finite temperature and baryon density. \cite{McLerran:2007qj},\cite{Hidaka:2008yy},\cite{Glozman:2007tv},\cite{Glozman:2008ja},\cite{Glozman:2008kn},\cite{McLerran:2008ua}  This matter exists at temperatures below that typical of de-confinement, and at baryon chemical potential larger than the nucleon mass, but less
than a baryon chemical potential of order $\sqrt{N_c} M_N$.  $N_c$ is the number of quark colors.
A phase diagram showing the set of possible phases of matter now expected is shown in Fig. \ref{blaschke}
Quarkyonic matter is described in the lectures of
Pisarski, and I will only quote a few of its properties:  It is confining and approximately chirally 
symmetric.  It is one of three distinct phases.  One phase is confined mesonic matter which has
approximately zero baryon number density.  Another is the Quark Gluon Plasma which is de-confined and has finite baryon number density. 
Quarkyonic matter has finite baryon number density, which may
be parametrically large compared to $N_c \Lambda_{QCD}^3$  It is confining: gluons are confined into glueballs,
and antiquarks into mesons in this phase.   

The existence of such matter is compelling in the large $N_c$ limit of QCD. The new phase is called Quarkyonic since because it is confined and therefore  it is composed of baryons, but also its  density can become so large that the correct way to describe its degrees of freedom, at least for those degrees of freedom deep within the Fermi sea, is through quarks. The phase diagram of nuclear matter, based on a detailed but model dependent
computation is shown in Fig. \ref{mclerran_pd}
\begin{figure}[!htb]
\begin{center}
  \mbox{{\epsfig{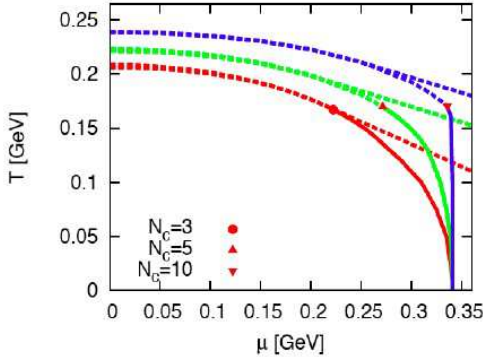}}
           }
   \end{center}
\caption[*]{\small The phase diagram for nuclear matter.  The confined phase is the enclosed region in the lower left hand part of the figure.  Quarkyonic matter is the region to the right of the solid line
that intersects the quark Fermi energy, $\mu_Q = \mu_B/N_c$ of $M_N/N_c \sim 0.35~ GeV$
at temperatures less than that of deconfinement. }
     \label{quarkyonic}
\end{figure}

We should ask: How does nuclear matter fit into this phase diagram?  First let us review the observed properties of nuclear matter:
\begin{itemize}

\item{ The baryon number density of nuclear matter is about $0.15~baryons/fm^3 \sim .1 -.2 ~N_c
~quarks/fm^3$.  This is about an order of magnitude smaller than $\Lambda_{QCD}^3$}

\item{The energy density of nuclear matter, including effects of the nucleon mass is of order
$.14 ~ GeV/fm^3$}

\item{The binding energy of nuclear matter is small, $\epsilon_B/N_B \sim 15~MeV$}

\item{Nuclear matter appears to be in a liquid phase.}

\item{Chiral symmetry appears to remain broken in nuclear matter.}

\end{itemize}
While the first two items above are problematic, the third is a significant problem for large $N_c$ physics.  This is because in the large $N_c$ limit baryon interactions are strong, and the expected binding energy of strongly interacting matter should be of order the nucleon mass.  This is verified in explicit computations, where it is shown that a Skyrme crystal is formed in the large $N_c$ limit.\cite{skyrme_crystal}  Not only is the binding energy of nuclear matter very small, it behaves like a liquid, not a solid.  It would appear that nuclear matter has little to do with quarkyonic matter.

The weakness of the binding energy of nuclear matter arises because the nuclear force at long and intermediate distance
does not generate potential energies of order the nucleon mass.  
If the large $N_c$ phenomenology naively applied was correct, then one would expect strong potentials,
with typical energy scales of order the nucleon mass.  It is sometimes argued that this is a result of accidental cancellations between the forces generated by various meson exchanges, but this cannot explain why the nucleon-nucleon potential remains small at such distances for a variety of  parameters, such as the pion mass and lattice spacing, in lattice gauge theory computations.\cite{Ishii:2006ec}

\section{Large $N_C$ and the Sigma Model}

To understand the origin of the problem of the large $N_c$ limit and the strength of nucleon potentials, consider  the linear sigma model.  The action for this theory is
\be
  S  & =  & \int~d^4x~ \left\{ \bar{\Psi}{1 \over i} \gamma \cdot \partial \psi \nonumber \right. \\
  & &  ~~~~~~ + g_{\pi NN} \bar{\psi}
  \left( \sigma + i \pi \cdot \tau \gamma_5 \right) \psi  \nonumber \\ & & \left.+ {1 \over 2} (\partial \sigma )^2+ {1 \over 2} (\partial \pi )^2
  + V(\sigma^2 + \pi^2) \right\}
 \ee
In this equation,
\be
 V(\sigma) = -\mu^2 \sigma^2/2 +\lambda \sigma^4/4
\ee
In the large $N_c$ limit, we expect that $\lambda \sim 1/N_c$, since mesons are weakly interacting, and that $g_{\pi NN} \sim \sqrt{N_c}$.  If we minimize the potential $V$, we get a vacuum expectation value for the Higgs field that is $\sigma_O = \mu/ \sqrt{\lambda} \sim \sqrt{N}$, so that $M_N \sim g_{\pi NN} \sigma_0 \sim N_c\mu$.  This is consistent with expectations if we choose the sigma meson
mass to be of order $\Lambda_{QCD}$.

It naively appears that the pion-nucleon coupling is strong, but this is not really the case.  Because of the $\gamma_5$ in the action above, in matrix elements involving the pion emission, there will be a factor of $1/M_N$, so that effectively, the interaction strength is of order $1/\sqrt{N_c}$  Such pion exchanges would indeed generate potentials that at the QCD distance scale were of order $1/N_c$.  This can be seen explicitly to all orders of pion emission using Weinberg's methods, and find that each pion emission is suppressed by $1/ \sqrt{N_c}$ \cite{Weinberg:1966fm}

Kaiser, Fritsch and Weise argued that using a low energy pion nucleon theory combined together with a repulsive hard core interaction, that the binding energy of nuclear matter generates a series in powers
of the Fermi momentum,\cite{Kaiser:2001jx}
\be
  \epsilon_{binding}/N_B \sim k_f^2/M - \alpha g^2_{\pi NN} k_F^3 /M_N^2  & & \nonumber \\
  + \beta g^4_{\pi NN} k_F^4/M_N^3 & &
\ee 
This expression is parametrically of order $1/N_c$ if we take the pion nucleon coupling to be of order 
$g_{\pi NN} \sim \sqrt{N_c}$.  In this derivation, little is used of the properties of the hard core other than it excludes the nucleon in some region.  Such a hard core could be of strength $N_c$.

How does the contradiction arise between the physics of the $\sigma$ model and that of conventional large $N_c$ dynamics?  If we generalize the linear sigma model to the non-linear one, which should be valid for long distance interactions $r >> \mu$, then using
\be
	& U = e^{i \pi \cdot \tau \gamma_5/f_\pi}
\ee
the Fermion interaction term becomes $M_N \bar{\psi} 	U \psi$.  We can rotate this into a pure mass term by rotating  the nucleon fields by   the "square root" of $U$, $ 1 = U^{-1/2} U U^{-1/2}$.   This rotation generates an interaction term
\be
  &  \bar{\psi} {1 \over i} \gamma^\mu U^{1/2} \partial_\mu U^{-1/2} \psi
\ee     
The parameter $1/f_{\pi} \sim g_{\pi
NN}/2M_N \sim 1/\sqrt{N}$ in this expression shows that for each pion emitted from a nucleon, there is a suppression of $1/\sqrt{N_c}$.
To first order in the pion field,
the interaction term is of order $(\partial^\mu \pi) \bar{\psi} \gamma_\mu \gamma_5 \psi$.  The derivative of the pion field couples to the axial vector current, and we are taking this coupling as one within the non-linear sigma model.  The axial vector coupling is in fact modified due to particle interactions,
and the axial vector coupling is introduced.  A Goldberger-Trieman relation is maintained,
\be
	 g_A/f_{\pi} = g_{\pi NN}/2 M_N
\ee
The problem with the large $N_c$ counting arises because in the non-relativistic quark model and in the Skyrme model\cite{Adkins:1983ya}
\be
   &   g_A \sim N_c 
\ee      
In the Skyrme model and conventional large $N_c$ counting, $g_{\pi NN} \sim N_c^{3/2} $!
This is also an issue for magnetic moments computed in large $N_c$.  This has a consequence
that the scalar excitation corresponding to the $\sigma $ meson, be it a resonance or a broad
enhancement, would naively be expected to interact very strongly with the nucleon, in a way inconsistent with naive large $N_c$ counting.

\section{Issues Concerning Large $N_c$ and the Skyrme Model}

The Skyrme model posits a non-linear sigma model for the pion with a fourth derivative term chosen because it is the only one of second order in time derivatives.  The action itself is formally of order $N_c$. It predicts a nucleon of size $1/\Lambda_{QCD}$ with a mass of order $N_c \Lambda_{QCD}$.
The fourth order term coupling $1/e^2 \sim N_c$, but numerically it is of order $1/125$.  It predicts a $\Delta$ nucleon split from the proton by a mass difference of order $\Lambda_{QCD}/N_c$.  The axial coupling is predicted to be approximately $g_A \sim (N_c+2)/3$, and magnetic momenta are also proportional to $N_c$.  

The Skyrme model has resisted derivation from first principles in QCD.  Attempts to do so typically generate not only a Skyrme term for non-linear interactions but several other terms which are not of second order in time derivatives.  These terms are of the wrong sign to stabilize the Skyrmion, so it may collapse to a smaller size scale.\cite{Aitchison:1984ys}

If we take the non-linear sigma model seriously, $g_A = 1$, and ask what is the size we expect for a nucleon,
the only scale in the problem is $1/ f_{\pi}$.  This is of order $1/\sqrt{N_c}$.  If one assumes that the nucleon has a much smaller size than $1/\Lambda_{QCD}$, then the non-linearities of the Skyrme model do not become important until distance scales which are small compared to $\Lambda_{QCD}$.
Before we would be able to explore such scales, higher derivative linear terms become important for
the Skyrme action.  Although this might prove a way out of these unpleasant features of  the Skyrme
mode, it also would have the bad feature that the naive $N_c$ expectations for the size and
mass of the nucleon are no longer obvious, and probably not true.

\section{Provocative Questions}

I want to end this presentation with a few questions for discussion.  These are only a small subset of the questions that I have.  To some of you the answers may seem obvious, and it is silly to ask such questions.  Nevertheless, for me and some of my colleagues, the observations noted above are troubling.  

\begin{itemize}
\item{Is the Skyrme model true and the binding energy of nuclear matter an accident?}

\item{Is the scalar channel of nucleon-nucleon interactions  super strong?}

\item{Is the implication within the Skyrme model that a nucleon is a pionic soliton not correct?}

\item{Is there something wrong with the parametric assignment of coupling strengths and masses  in powers of $N_c$ as is assumed in the Skyrme model?}

\item{Is the Skyrme model correct, but computations of the potential within the Skyrme model missing some essential physics?}

\end{itemize}

\section{Summary}

The limit of large $N_c$ the conventional description of nucleons presents apparent paradox.  It may be that the implications of
the Skyrme model are true, and accidently live in world at $N_c = 3$, where many of the properties of baryon do not reflect the large $N_c$ limit.  It might also be true that we are missing something fundamental in our understanding of nucleons.

\section{Acknowledgements}
 I thank the organizers of Fundamental Challenges of QCD, and in particular Leonya Glozman
 who made this meeting so pleasant.  I also thank my colleagues Jean-Paul Blaizot, Matiej Nowak and Rob Pisarski with whom I have had many heated conversations on this subject,
with whom many of these questions were formulated, and with whom research is in progress.  I particularly enjoyed the animated discourses with  Tom Cohen and 
Mitya Diakanov.
This manuscript has been authorized under Contract No. DE-AC02-98H10886 with
the U. S. Department of Energy.

\end{document}